\documentclass{article}%
\usepackage{amsmath}
\usepackage{amsfonts}
\usepackage{amssymb}
\usepackage{graphicx}%
\begin{document}
{\LARGE On the Raman shift in nanosized crystals}%
\[
\]
M. Salis$^{a,\ast}$, P. C. Ricci$^{a}$, A. Anedda$^{a,b}$

$^{a}$\textit{Department of Physics, University of Cagliari, s.p. 8 Km 0.7,
09042 - Monserrato, Cagliari, Italy}

$^{b}$\textit{LIMINA Laboratory, University of Cagliari, s.p. 8 Km 0.7, 09042
- Monserrato, Cagliari, Italy}

*Electronic address: masalis@unica.it%
\[
\]
\textbf{Abstract}. An analytical form of the Raman shift dependence on size of
nanocrystals is presented. Based on the hard confinement model, it works in
those cases where the average phonon curve shows a quadratic dependence on the
phonon quasi-momentum in the range of interest.

PACS : 78.67.Bf, 78.30.-j, 78.30.Am, 36.20.Ng
\[
\]
Raman spectroscopy is an important tool in gathering information about
molecular and crystal vibrational properties. Raman spectra can show
significant changes when sizes of investigated crystals go down to the
nanometer scale. Strain, stress and non-stoichiometry can be invoked to
explain these changes $[1,2]$. However, many published works have shown that
in most cases the main cause of spectra changes is to be ascribed to the
phonon confinement (PC) effect $[3-9]$. Actually, the confinement of the
$q_{0}=0$ phonon wave function in nanometric-sized crystals makes accessible
to Raman investigation a significant portion of the Brillouin Zone (BZ), whose
extension increases as the crystal size decreases $[3-5]$. Besides some
specific cases $[5,10,11]$, a direct relation, even when approximated, which
explicitly shows the connection between Raman shift and crystal size,
depending on suitable model parameters, is\ still \ not available. In this
letter we present a detailed numerical analysis of the hard confinement model
(HC) which yields an unexpectedly simple analytical form capable of closely
reproducing the model predictions. It uses the quadratic approximation of the
crystal phonon curve $\omega(q)$ which, in several practical cases, holds down
to a few nanometers of crystal sizes.

Formally, the PC effect can be explained by considering the modulation of the
phonon wave function in the infinite crystal with a suitable "weighting"
function $[4]$. On expansion by Fourier integrals, we can calculate the first
order Raman spectrum of a nano-sized crystal as
\begin{equation}
I(\omega)\propto\left[  n(\omega)+1\right]
%TCIMACRO{\dint \limits_{BZ}}%
%BeginExpansion
{\displaystyle\int\limits_{BZ}}
%EndExpansion
\frac{\left\vert C(\mathbf{q})\right\vert ^{2}}{\left[  \omega-\omega
(\mathbf{q})\right]  ^{2}+(\Gamma_{0}/2)^{2}}d\mathbf{q} \label{RamX0}%
\end{equation}
where $C(\mathbf{q})$ stand for Fourier coefficients of the $q_{0}=0$ phonon
wave function, $\omega(\mathbf{q)}$ for the phonon dispersion curve,
$\Gamma_{0}$ for the intrinsic Raman band linewidth and\textbf{ }$\left[
n(\omega)+1\right]  $\textbf{ }for the Bose-Einstein factor $[4]$; the latter
being usually disregarded for lineshape characterization or nanoparticle
sizing. Some simplifications are required in order to handle eq (\ref{RamX0}).
A drastic but commonly accepted assumption considers an isotropic dispersion
in a spherical BZ $[4]$. In this case, the function $\omega(q)$ represents an
average dispersion curve $[6]$. As concerns phonon confinement , the HC model
assumes a gaussian weighting function for spherical nanocrystals. Accordingly,
by disregarding unessential factors, eq (\ref{RamX0}) can be rewritten as
$[6,8,15]$%

\begin{equation}
I(\omega)=%
%TCIMACRO{\dint \limits_{0}^{\pi/a}}%
%BeginExpansion
{\displaystyle\int\limits_{0}^{\pi/a}}
%EndExpansion
\frac{\exp(-q^{2}L^{2}/16~\pi^{2})~q^{2}}{\left[  \omega-\omega(q)\right]
^{2}+(\Gamma_{0}/2)^{2}}dq \label{RamX1}%
\end{equation}
where $L$ stands for the particle size and $a$ for the lattice parameter. The
latter equation marks the start line of the present work.

If the quadratic approximation holds the phonon curve $\omega(q)$ can be
represented as%
\begin{equation}
\omega(q)=\omega_{0}+Aq^{2}a^{2}/2 \label{RamXQuad}%
\end{equation}
where $A$ stands for a suitable curve parameter. Equation (\ref{RamXQuad})
provides an univocal correspondence between peak frequency $\widetilde{\omega
}$ and an effective quasi-momentum $\widetilde{q}$ which allows for
$\widetilde{\omega}=\omega(\widetilde{q})$. Thus, we should search for an
expression of $\widetilde{\omega}$ by investigating the proper definition of
$\widetilde{q}$.

For convenience , let us define the function%

\begin{equation}
G(q,L)=q^{2}\exp(-q^{2}L^{2}/16~\pi^{2}) \label{RamX2}%
\end{equation}
so that, $I\left(  \omega\right)  =%
%TCIMACRO{\dint \limits_{0}^{\pi/a}}%
%BeginExpansion
{\displaystyle\int\limits_{0}^{\pi/a}}
%EndExpansion
G(q,L)/\left\{  \left[  \widetilde{\omega}-\omega(q)\right]  ^{2}+(\Gamma
_{0}/2)^{2}\right\}  dq$ . By taking into account that at the maximum of
$I(\omega)$ the condition $\left[  d~I(\omega)/d\omega\right]  _{\omega
=\widetilde{\omega}}=0$ holds, we obtain the exact equation
\begin{equation}%
%TCIMACRO{\dint \limits_{0}^{\pi/a}}%
%BeginExpansion
{\displaystyle\int\limits_{0}^{\pi/a}}
%EndExpansion
\frac{~G(q,L)\left[  \widetilde{\omega}-\omega(q)\right]  }{\left\{  \left[
\widetilde{\omega}-\omega(q)\right]  ^{2}+(\Gamma_{0}/2)^{2}\right\}  ^{2}%
}dq=0 \label{RamX3}%
\end{equation}
Of course, in the framework of the HC model the latter equation is general,
that is, not related to a special law for $\omega(q)$. Thus, for a given
$\Gamma_{0}$ and $\omega(q)$, we can define the distribution-like function
\begin{equation}
F_{\widetilde{\omega}}(q,L)=\frac{~G(q,L)}{\left\{  \left[  \widetilde{\omega
}-\omega(q)\right]  ^{2}+(\Gamma_{0}/2)^{2}\right\}  ^{2}}~/~%
%TCIMACRO{\dint \limits_{0}^{\pi/a}}%
%BeginExpansion
{\displaystyle\int\limits_{0}^{\pi/a}}
%EndExpansion
\frac{G(q,L)}{\left\{  \left[  \widetilde{\omega}-\omega(q)\right]
^{2}+(\Gamma_{0}/2)^{2}\right\}  ^{2}}dq \label{RamX4}%
\end{equation}
so that
\begin{equation}
\widetilde{\omega}=%
%TCIMACRO{\dint \limits_{0}^{\pi/a}}%
%BeginExpansion
{\displaystyle\int\limits_{0}^{\pi/a}}
%EndExpansion
F_{\widetilde{\omega}}(q,L)\omega(q)dq \label{RamX5}%
\end{equation}
Now, by inserting form (\ref{RamXQuad}) into eq. (\ref{RamX5}) we can make
explicit the meaning of the effective quasi-momentum $\widetilde{q}$ , that is,%

\begin{equation}
\widetilde{\omega}=\omega_{0}+\frac{A}{2}a^{2}%
%TCIMACRO{\dint \limits_{0}^{\pi/a}}%
%BeginExpansion
{\displaystyle\int\limits_{0}^{\pi/a}}
%EndExpansion
F_{\widetilde{\omega}}(q,L)q^{2}dq=\omega_{0}+\frac{A}{2}a^{2}\widetilde
{q}^{2} \label{RamX6}%
\end{equation}

The analytical resolution of eq. (\ref{RamX3}) is quite concealed and its
reduction to a differential equation, even if simplified, is not helpful for
the purpose. However, for the sake of discussion, it is worth while to handle
eq. (\ref{RamX3}) in a more explicit form. By taking into account that
$G(\pi/a,L)\approx0$ ( $L\gg a$) and after integrations by parts, derivations
with respect to $L$ and by disregarding small terms we obtain
\begin{equation}
\Theta(L)=\frac{\Lambda(L)}{9\left[  1-\Theta(L)\right]  ^{2}+\Lambda
(L)}\left[  2+L\frac{d\Theta(L)}{dL}\right]  \label{RamX7}%
\end{equation}
where%
\begin{equation}
\Theta(L)=1-\frac{L^{2}\widetilde{q}^{2}}{24\pi^{2}} \label{RamX8}%
\end{equation}
and%

\begin{equation}
\Lambda(L)=%
%TCIMACRO{\dint \limits_{0}^{\pi/\widetilde{q}a}}%
%BeginExpansion
{\displaystyle\int\limits_{0}^{\pi/\widetilde{q}a}}
%EndExpansion
F_{\widetilde{\omega}}(\xi,L)\frac{\left(  1-\xi^{2}\right)  /\xi^{2}%
}{\left\{  \left(  1-\xi^{2}\right)  ^{2}+(1/\widetilde{q}ax)^{4}\right\}
}d\xi\label{RamX9}%
\end{equation}
with $x=\left(  \left\vert A\right\vert /\Gamma_{0}\right)  ^{1/2}$. Due to
the gaussian factor present in $F_{\widetilde{\omega}}(\xi,L)$ , if $L$ is
sufficiently large with respect to $a$ , the upper integration limit can be
considered as infinite so that $\Lambda$\ could be conveniently presented as a
function on $\eta=L/xa$ rather that on $L$. Thus, to avoid confusion, further
on we use special symbols \ $\Theta_{S}(\eta)=\Theta(L)$ and $\Lambda_{S}%
(\eta)=\Lambda(L)$ even when small size are considered.

It is evident from eq. (\ref{RamX9}) that as $L$ increases $\Lambda_{S}(\eta)$
decreases because $\widetilde{q}\rightarrow0$. Thus, for large $L$ the
effective quasi-momentum decreases as $\widetilde{q}^{2}\approx24\pi^{2}%
/L^{2}$ and $\Lambda_{S}(\eta)$, as well as $\Theta_{S}(\eta)$, decreases as
$\sim1/\eta^{4}$. In the range of small sizes, eq. (\ref{RamX7}) cannot be
used. However, from eq. (\ref{RamX3}) we expect that for vanishing crystal
sizes $\Theta_{S}(\eta)\rightarrow1$ because of $\widetilde{q}\leq\pi/a$. In
the limiting case of $\widetilde{q}\rightarrow$ $\pi/a$ we expect $\Theta
_{S}(\eta)\sim\exp(-x^{2}\eta^{2}/24)$. A more detailed description of
$\Theta_{S}(\eta)$ is obtained from the numerical calculation of
$\widetilde{q}$. Results are shown in the inset of Fig.1 as obtained for
different $x$ values. For comparison, segments of parabola approaching the
curves $-ln\left[  \Theta_{S}(\eta)\right]  $ are shown as well. Precisely,
parabola have equations $f(\eta)=mx^{2}\eta^{2}/24$\ where $m=0.65$, $m=0.8$,
$m=1$ and $m=1$ for $x=0.5$, $x=1$, $x=2$ and $x=3$, respectively.

Usually $A\gtrsim\Gamma_{0}$ so that our interest will be focused on the
$x\gtrsim1$ curves. For $\eta>7$ the latter curves are independent of $x$ and
quickly approach the straight line $g(\eta)=\eta/12$, maintaining this trend
up to $\eta\approx60$ where $\Theta_{S}(\eta)<<1$ (Fig.1). On account of eq.
(\ref{RamX8}), we should not be concerned about the deviation of $-ln\left[
\Theta_{S}(\eta)\right]  $ from the linear law since for $\eta\gtrsim60$ we
have $\widetilde{q}^{2}\approx24\pi^{2}/L^{2}$. Thus we can be confident of
the approximation
\begin{equation}
\widetilde{q}_{ap}^{2}=\frac{24\pi^{2}}{L^{2}}\left[  1-\exp(-L/12ax)\right]
\label{RamApprox}%
\end{equation}
and, consequently,
\begin{equation}
\omega_{ap}(\widetilde{q})=\omega_{0}+12\pi^{2}\frac{Aa^{2}}{L^{2}}\left[
1-\exp(-L/12ax)\right]  \label{RamX10}%
\end{equation}
in the size range $L\gtrsim10xa$ (fig.1). Of course, when using eq.
(\ref{RamX10}) in practical cases, we should make sure that the quadratic
approximation holds down to the size dealt with.

Now, as an example, let us consider the case of Raman shift in titanate
($TiO_{2}$) nanocrystals for which was proposed the average phonon curve
$[12,13]$
\begin{equation}
\omega(q)=\omega_{0}+A\left[  1-\cos(qa)\right]  \label{RamX11}%
\end{equation}
where $\omega_{0}=144~cm^{-1}$ , $A=20~cm^{-1}$. and $\ a=0.376~nm$, the
linewidth being $\Gamma_{0}=7~cm^{-1}$ so that $x=1.7$. Fig. 2 \ shows the
deviations of Raman shifts from $\omega_{0}$ as calculated from eq.
(\ref{RamX1}) with $\omega(q)$ given in eq. (\ref{RamX11}) (triangles), as
calculated from eq. (\ref{RamX6}) (circles) and as calculated from eq.
(\ref{RamX10}) (dashed curve). As expected from Fig.1, in the range of small
sizes, the latter equations underestimates the deviations of shift. However,
departure from the exact curve remains relatively small down to 5 nm where the
PC model is believed to become unreliable~$[6,12,13]$.

A further example is given by the case of silicon nanocrystals whose phonon
curve, established by neutron scattering measurements $[14]$, is
\[
\omega(q)=\left[  B_{1}+B_{2}\cos(qa/4)\right]  ^{1/2}%
\]
with $a=0.5483nm$, $B_{1}=1.714~10^{5}~cm^{-2}$ , $B_{2}=10^{5}cm^{-2}$,
corresponding to $\omega_{0}=520.5~cm^{-1}$ and, in the quadratic
approximation, $A=-6~cm^{-1}$; the linewidth is $\Gamma_{0}=3.6~cm^{-1}$ so
that $x=1.3$. In this case deviations have a reversed sign. Departures from
the exact curve have roughly the same relative values found in the previous
case. It can be noted that because of the factor 1/4 in the cosine argument,
the quadratic approximation appears to hold in the whole range considered.

As a final consideration we note from Fig. 1 that one could improve the
approximation of curve (\ref{RamX10}) (cases x$\gtrsim1$) by introducing a
suitable factor to $\Theta_{S}(\eta)$. In this connection, we found that
$-ln\left[  \Theta_{S}(\eta)\right]  =0.3\eta\exp(-\eta/2)+\eta/12$ improves
the agreement down to $\eta\sim6$. However, in our opinion, this correction
does not recompense the loss of simplicity in eq. (\ref{RamX10}).

In conclusion, we have shown that in the framework of the PC effect as
described by the hard confinement model, the Raman shift in nanocrystals can
be calculated, with good approximation, by means of a simple analytical form
in the size range where the quadratic approximation of the average phonon
curve holds. The goal is that the dependence of shifts on model parameters as
well as on particle sizes is immediately recognizable.%

\[
\]
\textbf{References}

[1] V. Swamy,A. Kuznetsov, L. S. Dubrovinsky, R. A. Caruso, D. G. Shchukin and
B. C. Muddle, Phys. Rev. B \textbf{71}, 184302 (2005).

[2] T. Mazza, E. Barborini, P. Piseri, P Milani, D. Cattaneo, A. Li Bassi, C.
E. Bottani and C. Ducati, Phys. Rev. B \textbf{75}, 045416 (2007).

[3] H. Richter and Z. P. Wang, L. Ley, Solid State Comm. \textbf{39}, 625 (1981).

[4]I. H. Campbell and P. M. Fauchet, Solid State Comm. \textbf{58,} 739 (1986).

[5] G. Faraci,S. Gibilisco, P. Russo, A. R. Pennisi.and La Rosa, Phys. Rev. B
\textbf{73}, 033307 (2006).

[6] A. Dieguez, A Romano-Rodriguez, a: Vil\`{a} and J. R. Morante, J. App.
Phys \textbf{90}, 1550 (2001).

[7] R. Carles, A. Mlayah, M. Amjoud, A. Reynes and R. Morancho, Jpn. J. Appl.
Phys. \textbf{31}, 3511 (1992).

[8] P. C. Ricci, A. Casu, G. De Giudici, P. Scardi and A. Anedda, Chem. Phys.
Lett. \textbf{444}, 135 (2007).

[9] S. N. Klimin, V. M. Fomin, J. T. Devreese and D. Bimberg, Phys. Rev. B,
\textbf{77} 045307 (2008).

[10] V. Paillard, P. Puech, M. A. Laguna, R. Carles, B. Khon and F. Huisken,
J. App. Phys. \textbf{86}, 1921 (1999).

[11] J. Zi, K. Zhang and X. Xie Phys. Rev B \textbf{55}, 9263 (1997).

[12] D. Bersani,P.P. Lottici and X. Z. Ding, Appl. Phys. Lett. \textbf{72}, 73 (1999).

[13] W. F. Zhang, Y. L. He, M. S. Zhang and Z. Yin, Q. Chen, J. Phys. D: Appl.
Phys, \textbf{33}, 912 (2000).

[14] H. Xia, Y. L. He, L. C. Huang, W. Zhang, X. N. Liu, X. K. Zhang and Feng
J. Appl. Phys. \textbf{78}, 6705 (1995).%

\[
\]
\textbf{Captions of Figures}%
\[
\]

Fig.1 Negative logarithm of the function $\Theta_{S}(\eta)$(eq. 10). Inset,
dependence of $-ln\left[  \Theta_{S}(\eta)\right]  $ on x for small $\eta$: a,
x=0.5; b, x=1; c, x=2; d, x=3. Dashed curves represent parabola with equations
$f\left(  \eta\right)  =mx^{2}\eta^{2}$: a, m=0.6; b, m=0.8; c, m=1; d, m=1.%

\[
\]

Fig. 2. Deviations of Raman shifts in nanocrystals from the Raman shift in the
infinite crystal. Curves are referred to the source equations. The upper
curves have been calculated for the Titanate and the lower curves for the Silicon.

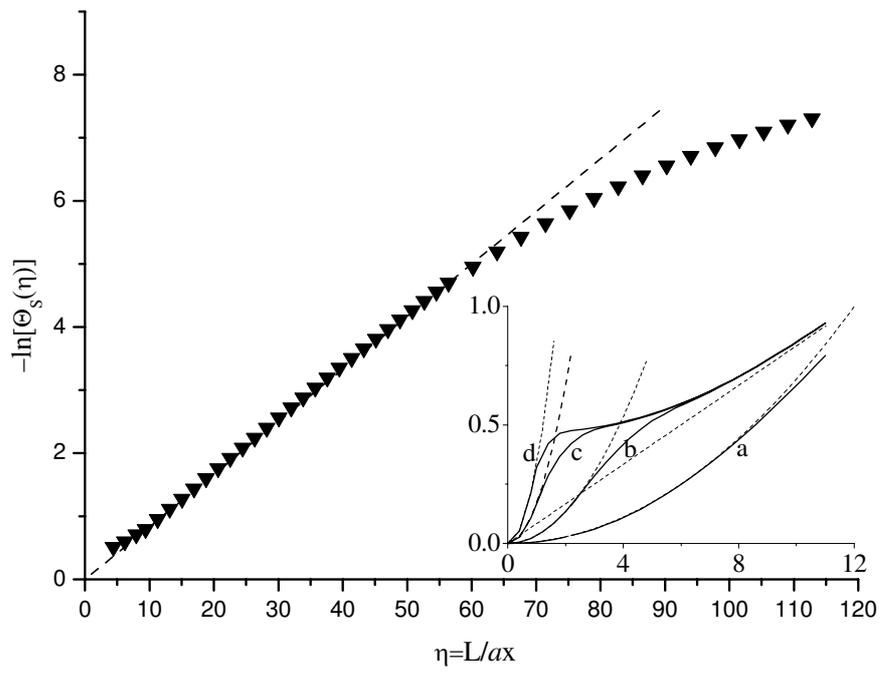
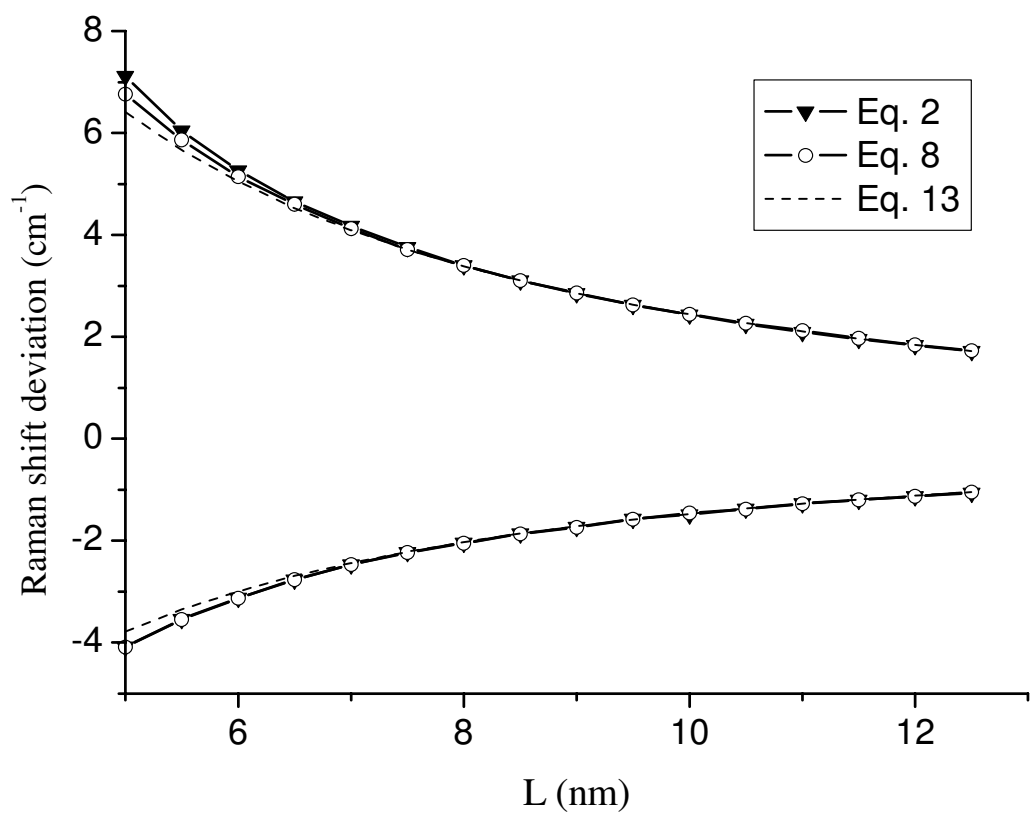
\end{document}